\magnification=\magstep1
\font\sc=cmcsc10 scaled 1200
\def\pcap#1{ \sc {{#1}}}
\def\etal{{et al.\ }}
\def\sun{\odot}
\def\ref{\par \noindent \hangafter=1 \hangindent 22.76pt}
\def\s{\hbox{s}}
\def\yr{\hbox{yr}}
\def\mathnew{\mathsurround=0pt}
\def\simov#1#2{\lower .5pt\vbox{\baselineskip0pt \lineskip-.5pt
        \ialign{$\mathnew#1\hfil##\hfil$\crcr#2\crcr\sim\crcr}}}
\def\simgreat{\mathrel{\mathpalette\simov >}}
\def\simless{\mathrel{\mathpalette\simov <}}

\centerline{GRO J1744-28: LAST GASPS OF A DYING LOW-MASS X-RAY BINARY}

\centerline{\pcap D. Q. Lamb, M. Coleman Miller}

\centerline{Department of Astronomy \& Astrophysics, University of Chicago}

\centerline{5640 S. Ellis Avenue, Chicago, IL 60637}

\centerline{\pcap and R. E. Taam}

\centerline{Department of Physics \& Astronomy, Northwestern University,
Evanston, IL 60208}
\medskip

\centerline{ABSTRACT}

We argue that the bursting, transient X-ray source GRO J1744-28 is a
binary consisting of a very low-mass ($M \approx 0.2 M_\sun$), highly
evolved giant star that is transferring mass by Roche lobe overflow
onto a high-mass ($M \approx 1.8 M_\sun$) neutron star.  We explore a
picture in which the bursts are due to thermonuclear flashes in matter
that has accreted onto the neutron star.  We attribute the unusually
hard spectra of the bursts and the high burst rate compared to normal
Type I X-ray burst sources -- indeed the existence of unstable nuclear
burning itself -- to the unusually strong surface magnetic field, $ B_s
\approx 1 \times 10^{13}$ G, of the neutron star.

\ref {\it Subject headings:} Accretion --- Magnetic Fields ---
Stars: Binary: Close --- Stars:  Evolution --- Stars: Neutron ---
X-rays: Bursts
\medskip

\centerline{I. INTRODUCTION}
\medskip

Fishman \etal (1995) have reported the discovery of a rapidly bursting,
transient X-ray source, GRO J1744-28, in the direction of the galactic
center.  The bursts last 8 to $\bf >$ 30 sec, (Fishman \etal 1995) and
have a characteristic spectral energy $\langle E \rangle \approx 14$
keV (Briggs \etal 1996; Swank \etal 1996).  The rate of bursting has
declined from about 18 events per hour initially to about 1.6 events
per hour (Fishman \etal 1995; Fishman \etal 1996) as the persistent
emission has increased (Paciesas \etal 1996).  This bursting behavior
is unlike that previously seen from any transient X-ray or $\gamma$-ray
source.

Recent reports that both the X-ray bursts (Kouveliotou \etal 1996) and
the persistent X-ray emission (Finger \etal 1996a) from GRO J1744-28
exhibit coherent pulsations at a frequency $f = 2.14$ Hz demonstrates
conclusively that both come from the same source.  The lack of
absorption by cold interstellar gas of the X-ray spectra of both the
bursts and the persistent emission (Swank \etal 1996) suggests that the
system is closer than the galactic center, while the apparent absence
of an optical counterpart brighter than $m_g = 19$ or $m_r = 20.5$
(Miller \etal 1996) suggests that the system is at least 2 kpc
distant.  Assuming a value of 3 kpc for the distance and isotropic
emission, the current persistent X-ray flux $F_p \approx 2 \times
10^{-7}$ erg cm$^2$ s$^{-1}$ (Swank \etal 1996; Fishman \etal 1996)
implies a persistent X-ray luminosity $L_p \approx 2 \times 10^{38}$
erg s$^{-1}$.
\bigskip

\centerline{II. NATURE OF THE PULSATIONS}
\medskip

What can be said about the nature of the $f = 2.1$ Hz oscillations seen
in the burst and persistent X-ray emission?  We associate the
corresponding period, $P \approx 0.5$ s, with the rotation period of
the strongly magnetic neutron star.  We attribute the substantial
amplitude (pulsed fraction $\approx 0.5$) of the pulsations in the
persistent X-ray emission to funneling of the accreted matter onto the
magnetic poles of the neutron star, where X-rays are produced when the
accreting matter strikes the stellar surface, as in the standard
picture of accretion-powered pulsars (see., e.g., M\'esz\'aros 1992).
Indeed, the X-ray spectrum of the persistent emission is similar to
that of accretion-powered pulsars (M\'esz\'aros 1992), with a
characteristic energy $\langle E \rangle \approx 14$ keV, and an
exponential falloff above this energy (Briggs \etal 1996; Swank \etal
1996).
 
Further support for this picture comes from the observed frequency
derivative, $8.6 \times 10^{-12}$ Hz s$^{-1}$ (Finger \etal 1996b),
which corresponds to a spinup rate $\dot{P}_{\rm obs} = -5.6 \times
10^{-5}$ s yr$^{-1}$ for the neutron star.  This value is in the range
expected if the persistent X-ray luminosity comes from accretion of
matter onto the neutron star, and the neutron star is being spun up by
the accretion torque (Ghosh \& Lamb 1979), $\dot{P} = - 8.3 \times
10^{-5} \mu_{30}^{2/7} (P_{0.5} L_{38}^{3/7})^2 \; \s \; \yr^{-1}$.
Here $\mu_{30}$ is the magnetic dipole moment $\mu = B_{\rm dipole}
R^3$ of the neutron star in units of $10^{30}$ gauss cm$^3$, $P_{0.5}$
is its rotation period in units of 0.5 s, and $L_{38}$ is its accretion
luminosity in units of $10^{38}$ erg s$^{-1}$.  The fact that the
neutron star is spinning up requires that it not be a fast rotator
(Ghosh and Lamb 1979), while the lower-than-expected value of the
spinup rate suggests that the rotation period of the neutron star lies
near its equilibrium spin period $P_{\rm eq}$.  These constraints imply
$\mu_{30} \approx 1$ and thus $B_{\rm dipole} \simless 1 \times 10^{12}$
G.

The current persistent X-ray luminosity $L_p$ requires a mass accretion
rate $\dot{M} \approx 1 \times 10^{-8} M_\sun$ yr$^{-1}$, which
corresponds to the Eddington accretion rate $\dot{M}_E$ for
accretion over the {\it entire surface} of a $M \approx 1.8 M_\sun$
neutron star.  However, the dipole magnetic field $B_{\rm dipole}
\approx 1 \times 10^{12}$ G implied by the observed spinup rate funnels
the accreting matter onto a region near the magnetic pole of the
neutron star with a characteristic area $A_{\rm acc} \sim \pi (R/r_A)
R^2$ corresponding to a fraction $f_{\rm acc} \sim 10^{-2}$ of the
stellar surface.  Consequently, the accretion rate per unit area
$\dot{\sigma} \approx 10^7$ g cm$^{-2} \approx 10^2 \dot{\sigma}_E$,
where $\dot{\sigma}_E$ is the mass accretion rate per unit area that
yields the Eddington flux at the surface of a neutron star.

We conjecture that the ability of GRO J1744-28 to accrete at this rate
is due to an unusually strong surface magnetic field $B_s \approx 1
\times 10^{13}$ G at the magnetic pole of the neutron star.  Such a
strong field funnels the flow onto a region near the magnetic pole of
the neutron star, thereby creating a highly non-spherical geometry in
which the accreting matter falls vertically onto the stellar surface
while the radiation produced escapes primarily laterally out the sides
of the accretion column.  Equally important, a field this strong
dramatically reduces the electron scattering cross section, which is
the dominant opacity in the accretion column, for radiation escaping
parallel and perpendicular to the magnetic field (see, e.g.,
M\'esz\'aros 1992).  In particular, the electron scattering cross
section becomes $\sigma_{e\gamma} \approx (\omega/ \omega_C)^2
\sigma_T$ for extraordinary mode photons traveling at large angles to
the magnetic field with energies $\hbar \omega \ll \hbar \omega_C$, the
cyclotron energy.  For photons with energy $\hbar \omega \approx
\langle E \rangle \approx 14$ keV, $\sigma_{e\gamma} \approx (14 \;
{\rm keV}/120 \; {\rm keV})^2 \sigma_T \approx 10^{-2} \sigma_T$.

Consequently, the critical luminosity, at which radiation force
balances gravity, becomes $L_c \approx 5 (\omega_C/ \omega) L_E \approx
10^2 (B_s/10^{13}$ G) $L_E$ for a plane parallel geometry (Miller
1995).  For the highly non-spherical (non-plane parallel) geometry of
the accretion column in GRO J1744-28, we expect an even larger value
for $L_c$.  Under these conditions, in which $L_p \approx L_c$, we
expect the accretion shock to be radiation-dominated (M\'esz\'aros
1992), and the resulting emission to escape in roughly the form of a
fan beam.  Such a fan beam provides a natural explanation of the
observed sinusoidal light curve of the persistent emission.

Similarly high accretion luminosities are seen in a number of
accretion-powered pulsars having high mass companion stars, such as
A0538-66, LMC X-4, and SMC X-1 (see, e.g., Stella, White, \& Rosner
1986).  These high accretion luminosities have never been
satisfactorily explained.  We conjecture that a very strong surface
magnetic field $B_s \simgreat 1 \times 10^{13}$ G may be the
explanation of the high luminosities in these sources as well.

There is a substantial difference between the dipole magnetic field
$B_{\rm dipole} \approx 1 \times 10^{12}$ G implied by the observed
spinup rate and the surface magnetic field $B_s \approx 1 \times
10^{13}$ needed in order to make possible the accretion rate per unit
area $\dot{\sigma} \approx 10^2 \dot{\sigma}_E$ that is required to
explain the persistent X-ray luminosity of these sources.  A similar
difference is seen in, e.g., Her X-1 between the value $B_{\rm dipole}
\approx 5 \times 10^{11}$ G implied by its spinup rate and the value
$B_s \approx 3 \times 10^{12}$ G indicated by the energy of the
cyclotron scattering feature in its X-ray spectrum (Ghosh \& Lamb
1979).  These differences suggest that the magnetic fields of the
neutron stars in these, and perhaps other, systems are not simple
dipoles.
\bigskip

\centerline{III.  EVOLUTIONARY STATUS OF THE BINARY}
\medskip

A timing analysis of the pulsations in the persistent X-ray emission
from GRO J1744-28 has revealed a sinusoidal variation in the phase of
the pulsations, implying that the X-ray source is a low-mass binary
with an orbital period $P_b = 11.76$ days (Finger \etal 1996b).  The
companion star must have a mass $0.2 M_\sun \simless M_c \simless 1.1
M_\sun$ (see below).  For an assumed neutron star mass $M_{ns} = 1.4 -
2.0 M_\sun$, the derived X-ray mass function $f(M) = 1.31 \times
10^{-4} M_\sun$ corresponds to companion star masses $M_c$ in the range
$0.2 - 1.1 M_\sun$ for inclination angles $i = 5^\circ - 25^\circ$.
However, it is likely that $M_c \approx 0.2 M_\sun$, since the small
values of $i$ required for larger values of $M_c$ are highly improbable
(see Table 1).

What can be said about the evolutionary status of the binary?  The
relatively short rotation period of the neutron star implies that it
has been accreting at a mean rate $\langle \dot{M}\rangle \approx 1
\times 10^{-8} M_\sun$ yr$^{-1}$ for at least the spinup timescale
$\tau_{\rm spinup} = -P/\dot{P} \approx 1 \times 10^4$ yrs, and most
likely for far longer.  The system may have gone undetected during the
past twenty years or so if the mass transfer rate during this time was
somewhat smaller than currently, and that consequently, the neutron
star was a fast rotator and did not accrete much matter (Illarionov \&
Sunyaev 1975).

A mass transfer rate as high as $\dot{M}_{\rm tr} \approx 1 \times
10^{-8} M_\sun$ yr$^{-1}$ lasting for more than $10^4$ yrs is strong
evidence that mass transfer in the binary is occurring via Roche lobe
overflow, and consequently that the low-mass secondary fills its Roche
lobe.  Further evidence that the secondary fills its Roche lobe comes
from the present low eccentricity $\epsilon < 0.026$ (90\% CL; Finger
\etal 1996b) of the binary.  This implies that the low-mass secondary
was once or is now comparable in size to the binary separation, in
order that tidal dissipation circularize the highly eccentric binary
resulting from the supernova explosion that produced the neutron star
(Zahn 1977).  The requirement that the secondary fill its Roche lobe
and the observed binary orbital period of 11.76 days implies that the
low-mass companion star is a highly evolved giant.

A binary consisting of a main sequence star and neutron star can lead
to the current binary configuration of GRO J1744-28 if the mass $M_{\rm
MS}$ of the main sequence star is in the range $\approx 0.8 - 1.1
M_\sun$ and the initial orbital period $P_b^0 \simless 10$ days.  The
main sequence star must have a mass $M_{\rm MS} \ge 0.8 M_\sun$ in
order for it to have evolved into a giant star within the age of the
universe, and a mass $M_{\rm MS} \le 1.1 M_\sun \approx (5/6) M_{ns}$
in order for the mass transfer process to be both conservative and
stable (Joss, Rappaport, \& Lewis 1987).  According to this scenario,
the main sequence star evolves into a lower giant-branch star,
consisting of a degenerate helium-rich core and an extended
hydrogen-rich envelope.  When the size of the giant reaches that of its
Roche lobe, mass transfer onto the neutron star commences, and
continues on the nuclear burning timescale of the giant (Webbink,
Rappaport, \& Savonije 1983; Taam 1983).  The resulting mean mass
transfer rate $\langle \dot{M}_{\rm tr} \rangle \approx 5 \times
10^{-9} (P_b^0/10 \; {\rm days}) \; M_\sun$ yr$^{-1}$ (Joss \&
Rappaport 1983; Paczy\'nski 1983; Savonije 1983), is somewhat smaller,
but not dissimilar to, the mass transfer rate implied by the current
persistent X-ray luminosity.  The initial orbital period $P_b^0$ must
be $\simless 10$ days since the binary separation, and hence the binary
orbital period, increases as the secondary star evolves.

The parameters of the current binary system can be calculated using the
radius--core mass relationship and the luminosity--core mass
relationship for lower-branch giant stars (Joss, Rappaport \& Lewis
1987).  We assume that, initially, the companion star had a mass $0.8
M_\sun \simless M_c \simless 1.1 M_\sun$, and a value $Z = 0.02$ for
its metallicity.  Table 1 shows the results for two cases: if Roche
lobe overflow is just beginning, and if Roche lobe overflow is just
ending.  We assume values for the neutron star mass $M_{ns}$ of $1.4
M_\sun$ and $2.0 M_\sun$ for these two cases; other values of $M_{ns}$
are allowable, provided that, following mass transfer, the final value
of $M_{ns}$ is less than the maximum stable mass of a neutron star.
Since we assume that the giant fills its Roche lobe, its properties do
not depend on $M_{ns}$.  Table 1 shows that it is highly likely that
$M_c \approx M_{\rm He} = 0.20 M_\sun$, and that the mass transfer
phase is just ending.  It is interesting to note that the radius--core
mass relationships for Population I and Population II lower-branch
giants cross at $M_{\rm He} \approx 0.20 - 0.23 M_\sun$ (Joss \etal
1987), so that our predictions for the parameters of the current binary
system are essentially independent of metallicity.

If our picture of GRO J1744-28 is correct, its current evolutionary
status is similar to that of other LMXBs in which the companion star is
a highly evolved, low-mass giant, such as Cir X-1, Cyg X-2, and the
X-ray transients Cen X-4 and Aql X-1 (Lewin, van Paradijs, \& Taam
1993).  All of these sources have been observed to produce Type I X-ray
bursts, which are thought to be due to thermonuclear flashes in matter
accreted onto the neutron star, but none of them exhibit the strong,
coherent pulsations in their persistent X-ray emission that are the
signature of a strongly magnetic neutron star.  The system GRO J1744-28
is thus unusual, in that only one other LMXB, 4U1626-67, which has an
orbital period $P_b \approx 42$ min and thus an entirely different
evolutionary history, exhibits coherent pulsations in its persistent
X-ray emission, implying the presence of a strongly magnetic neutron
star (Lewin \etal 1993).

The strong magnetic field of the neutron star in GRO J1744-28 implies
that the neutron star will {\it not} become a millisecond pulsar (cf.,
e.g, Joss \& Rappaport 1983), since the phase of mass transfer appears
to be nearly at an end.  Instead, it will eventually be observed as a
normal strong-field pulsar (unless it is spun down~(Illarionov \&
Sunyaev 1975, Ghosh \& Lamb 1979) to a period $P > 3$ sec as the mass
transfer from the giant companion tails off).  The existence of such a
strong magnetic field in an old neutron star has implications for
neutron star magnetic field evolution: it may contradict a
widely-discussed picture of how millisecond pulsars are formed, in
which the slowing down of the neutron star in an LMXB when mass
transfer first begins leads to a loss of magnetic flux from the neutron
star, reducing its dipole magnetic field from $B_{\rm dipole} \sim
10^{12}$ G to $B \sim 10^9$ G (Srinivasan \etal 1990).  Relatively
recent formation of the neutron star in GRO J1744-28 via
accretion-induced collapse, while potentially mitigating the question
of magnetic field decay, does not alter the conclusion that this system
will not become a millisecond pulsar, since the phase of mass transfer
appears to be nearly at an end.
\bigskip

\centerline{IV.  NATURE OF THE BURSTS}
\medskip

What might be the nature of the bursts produced by GRO J1744-28?  The
bursting behavior of the neutron star in this system is similar in some
ways, but not in others, to each of the three known classes of burst
sources: X-ray burst sources (Lewin \etal 1993) soft $\gamma$-ray
repeaters (Hurley 1996), and $\gamma$-ray burst sources (Fishman \&
Meegan 1995).  One possibility is that the bursts are due to
instabilities in the accretion flow onto the neutron star, as is
thought to be the explanation of the bursts seen in the singular source
called the ``Rapid Burster'' (Lewin \etal 1993).  Here we explore a
different picture in which the bursts are due to thermonuclear flashes
in matter that has accreted onto the neutron star; i.e., they are a
variant of Type I X-ray bursts.

A characteristic signature of Type I X-ray bursts is $\alpha \equiv
L_p/\langle L_b \rangle \approx 10 - 300$, where $\langle L_b \rangle$
is the time-averaged burst luminosity.  We can re-express $\alpha$ in
the distant-independent form $\alpha = (F_p/S_b r_b)$, where $S_b$ is
the burst fluence.  In early December 1995, the burst fluence $S_b$ was
$\approx 5 \times 10^{-7}$ erg cm$^{-2}$ in the 20-40 keV energy band,
and the burst rate $r_b$ was $\approx 5 \times 10^{-3}$ s$^{-1}$.
Because of the difficulty in deriving $F_p$ from Earth-occultation data
in the presence of other X-ray sources in the galactic center region
(particularly the bright X-ray source 1E1740.7-29, which lies only 0.8
degrees away from GRO J1744-28), the value of $F_p$ at that time is
uncertain (Paciesas \etal 1996).  However, $F_p$ was apparently
$\simless 2 \times 10^{-8}$ erg cm$^{-2}$ s$^{-1}$ in the 20-40 keV
energy band.  This yields a value $\alpha \simless 10$.  The unobserved
flux below 20 keV could affect these numbers significantly.  More
recently, $F_p \approx 2 \times 10^{-7}$ erg cm$^{-2}$ s$^{-1}$, $S_b
\approx 3 \times 10^{-6}$ erg cm$^{-2}$, and $r_b \approx 4 \times
10^{-4}$ s$^{-1}$ (Swank \etal 1996; Fishman \etal 1996), yielding
$\alpha \approx 200$.  Both the initial and the present $\alpha$-values
could differ substantially from the above estimates if the angular
distributions of the burst and the persistent emission differ.  Thus
the $\alpha$-values of the bursts from GRO J1744-28 lie in the range
$\simless$ 10 to $\approx$ 200, which is characteristic of Type I X-ray
bursts.
 
The temporal evolution of $\alpha$ from $\simless 10$ to $\approx 200$,
correlated with an increase in $L_p$, that is seen in GRO J1744-28 has
been observed in a number of Type I burst sources, most notably the
transient X-ray sources 0748-676 and 1658-298 (Lewin \etal 1993).
Within the framework of the thermonuclear flash model, such a change in
the $\alpha$-value of the bursts implies a transition from combined
hydrogen-helium flashes to nearly pure helium flashes (accompanied by
substantial steady helium burning; see Lewin \etal 1993).  The observed
timescale of a week or so for the transition corresponds to the
timescale for approach of the neutron star envelope to thermal
equilibrium.  At the high mass accretion rate per unit area
$\dot{\sigma} \approx 10^2 \dot{\sigma}_E$ we infer for GRO J1744-28,
thermal equilibrium will be accompanied by substantial steady hydrogen
burning due to electron captures deep in the envelope of the neutron
star (Taam \etal 1993, 1996).  If our picture of the bursts from GRO
J1744-28 is correct, we predict that the more recent bursts should be
shorter and have higher peak fluxes, as is typical of nearly pure
helium flashes compared to combined hydrogen-helium flashes.  The more
recent bursts may also be accompanied by mass loss via a wind during
the peak of the burst (Lewin \etal 1993).

Another signature of the thermonuclear flash model of X-ray bursts is a
rough correlation between the burst rate $r_b$ and the mass accretion
rate per unit area $\dot{\sigma}$.  There is substantial scatter in the
correlation between $r_b$ and $L_b$ in GRO J1744-28.  Although the
burst rate is much higher (which we attribute to the much higher
$\dot{\sigma}$ resulting from funneling of the accretion flow onto a
small polar cap area by the strong magnetic field of the neutron star;
see below), the general pattern seen in GRO J1744-28 is similar to that
seen in the Type I burst source 1636-536 (Lewin \etal 1993).  Indeed,
erratic burst behavior of this kind is characteristic of many Type I
burst sources, most notably 1735-44 (Lewin \etal 1993).  Within the
framework of the thermonuclear flash model of X-ray bursts, such
erratic behavior finds a natural explanation in terms of low
metallicity in the accreted matter (Taam \etal 1993), possibly because
the mass transferred from the Population II mass-losing secondary has
low metallicity or because spallation reactions in the accretion shock
at the neutron star surface deplete the abundance of heavy nuclei in
the accreted matter (Bildsten, Salpeter, \& Wasserman 1992).

Why do thermonuclear-powered bursts occur in GRO J1744-28, but
apparently not in other accretion-powered pulsars (Lewin \etal 1993)?
We conjecture that the explanation lies with the unusually strong
surface magnetic field, $B_s \approx 1 \times 10^{13}$ G, that we posit
for the neutron star in GRO J1744-28.

Such a strong magnetic field funnels the flow of accreting matter onto
a small area near the magnetic polar cap of the neutron star, as we
have discussed, resulting in accretion rates per unit area $\dot{\sigma}
\approx 10^7$ g cm$^{-2}$ s$^{-1}$.  Such a field is also strong enough
to confine the accreted matter in the envelope to the region of the
magnetic polar cap, since $P_{\rm gas} \simless (B_s^2/8\pi) \approx 4
\times10^{24}$ erg cm$^{-3}$.  The accreted matter thus forms a thin
disk of radius $r_{\rm acc} \approx 10^5$ cm and depth $d \sim 10^2
- 10^3$ cm, at the bottom of which nuclear burning occurs.

Normally, a high $\dot{\sigma}$ leads to a very high temperature $T$ in
the neutron star envelope, due to compressional heating (Fushiki \&
Lamb 1987).  The very high $T$ reduces the temperature sensitivity of
the nuclear reactions and stabilizes the nuclear burning.  However, as
we discussed earlier, a magnetic field as strong as that we posit for
GRO J1744-28 dramatically reduces the electron scattering cross
section, which is the dominant radiative opacity in the envelope, for
radiation escaping outward from the accreted matter.  The electron
scattering cross section becomes $\sigma_{e \gamma} \approx
(\omega/\omega_c)^2 \sigma_T \approx 10^{-2} \sigma_T$ for all photons
traveling along the magnetic field and for photons in the extraordinary
mode traveling at large angles to the field with energies $\hbar \omega
\ll \hbar \omega_C$.  

For $B_s \simgreat 3 \times 10^{12}$ G, the peak of the photon number
spectrum for a blackbody temperature $T \approx 1 \times 10^8$ K, a
temperature typical of the neutron star envelope, lies at an energy
$\hbar \omega \simless \hbar \omega_C$.  Under these conditions, the
enhanced radiative energy transport prevents the neutron star envelope
from reaching the very high $T$ otherwise expected for such a high
$\dot{\sigma}$.  It seems possible that the resulting physical
conditions of density and temperature are similar to those encountered
in normal Type I X-ray burst sources, and that consequently, hydrogen
and helium burning are highly unstable.

In contrast, for $B_s < 3 \times 10^{12}$ G, the peak of the photon
number spectrum for a blackbody temperature $T \approx 1 \times 10^8$ K
lies at an energy $\hbar \omega > \hbar \omega_C$, and the surface
magnetic field has little effect on the radiative opacity.  Under these
conditions, the temperature in the neutron star envelope reaches very
high values, due to compressional heating, which stabilizes the
nuclear burning.  Consequently, we do not expect Type I X-ray bursts to
occur in most accretion-powered pulsars.

Joss and Li (1980) found that a strong $B_s$, as well as a high
$\dot{\sigma}$, stabilizes nuclear burning, in a study that assumed
spherical symmetry and, most importantly, a very high neutron star core
temperature, $T_c \approx 4 \times 10^8$ K.  Under these conditions, the
enhanced radiative and conductive energy transport, due to the strong
magnetic field, brought the temperature in the burning shell $T_{\rm
shell}$ up to $T_c \approx 4 \times 10^8$ K, which stabilized the
nuclear burning.  Thus, the stabilizing effect they found is a result
of their particular assumption of a very high neutron star core
temperature $T_c$.  In general, and certainly in the case of a
transient X-ray source like GRO J174-28, we expect the envelope and the
core of the neutron star to be relatively cold (Fushiki \& Lamb 1987).
Then the calculations of Joss and Li (1980) do not apply, and a strong
magnetic field has a strongly de-stabilizing, rather than stabilizing,
effect on the nuclear burning.

Why is the X-ray spectrum of the bursts from GRO J1744-28 similar to
the spectrum of the persistent X-ray emission, which has a
characteristic spectral energy $\langle E \rangle \approx 14$ keV that
is typical of accretion-powered pulsars?  Since the surface magnetic
field is sufficient to confine the accreted matter to a thin disk lying
directly below the accretion column, and since the energy from the
thermonuclear flash is transported primarily outward, the photon flux
from the burst passes through the radiation-dominated accretion shock.
In this situation, we expect the radiation from the burst to be
upscattered by the accretion shock in the same way as is the radiation
generated by accretion (see Me\'sz\'aros 1992 and references
therein).  Consequently, we expect the X-ray bursts and the persistent
X-ray emission to have similar similar spectra.

Why is the burst rate $r_b$ in GRO J1744-28 so much higher than in
other Type I X-ray burst sources?  If, as we suggest above,
thermonuclear flashes occur in the envelope of the neutron star under
conditions similar to those of normal Type I X-ray bursts, $\rho
\approx 3 \times 10^5 - 3 \times 10^7$ g cm$^{-3}$, corresponding to
column densities $\sigma \approx 10^8 - 3 \times 10^{10}$ g cm$^{-2}$
(Lewin \etal 1993).  For the very high mass accretion rate per unit
area $\dot{\sigma} \approx 10^7$ g cm$^{-2}$ s$^{-1}$ appropriate to
GRO J1744-28, this implies burst rates $r_b \sim \dot{\sigma} / \sigma
\sim 3 \times 10^{-4} - 10^{-1} (\dot{\sigma} / 10^7$ g cm$^{-2}$
s$^{-1}$) ($\sigma/10^9 - 3 \times 10^{10}$ g cm$^{-2}$) s$^{-1}$.
This is just the range of burst rates that is observed.

Lastly, we remark that one of the few Type I X-ray bursts observed
from the X-ray transient Aql X-1 exhibited coherent oscillations of
modest amplitude (0.1 pulsed fraction) at $f = 7.6$ Hz, which have been
interpreted as due to rotation of the underlying magnetic neutron star
at a period $P = 0.13$ s (Schoelkopf \& Kelley 1991).  Thus GRO
J1744-28 may not be unique among Type I X-ray burst sources in having a
significant surface magnetic field.

Tests of the thermonuclear flash model of the bursts from GRO J1744-28
include the following:

(1) Observation of a cyclotron resonant scattering line in the spectrum
of GRO J1744-28 at an energy $E_c \approx 120 (B_s/10^{13} {\rm G})$
keV would provide strong support for the model; conversely, observation
of a cyclotron resonant scattering line at an energy significantly below
this would rule out the model.

(2) Confirmation that the $\alpha$-values of the bursts were $\approx
10$ initially would support the thermonuclear flash model, while a much
smaller upper limit on the initial value of $\alpha$ would pose
difficulties for the model.

(3) The luminous Type I bursts produced by GRO J1744-28 are capable of
disrupting the structure of the inner accretion disk, due to the
radiation force and/or ablation from the outflowing wind produced by
the bursts.  This could lead to a reduction in the mass accretion rate
$\dot{M}$, and therefore $L_p$, following the bursts, as observed
(Swank \etal 1996).  A decrease in $\dot{M}$, and therefore $L_p$,
before the bursts would be more difficult to understand within
framework of thermonuclear flash model.
\bigskip

\centerline{V. CONCLUSION}
\medskip

In conclusion, if the picture that we have developed of the newly
discovered bursting, transient X-ray source GRO J1744-28 is correct,
the evolutionary status of the binary system is similar to that of
other LMXBs with long orbital periods $P_b$, but the presence in it of
a neutron star with a very strong magnetic field is unusual.  We have
explored a picture in which the bursts are due to thermonuclear flashes
in the matter that has accreted onto the neutron star, and are
therefore a variant of Type I X-ray bursts.  We attribute the rapid
burst rate and the hard spectra of the X-ray bursts from GRO J1744-28
-- indeed the existence of unstable nuclear burning itself -- to the
unusually strong surface magnetic field of the neutron star.
\smallskip

This research was supported in part by NASA grant NAG 5-2868, NAGW
2526, and NASA contract NASW-4690.  MCM gratefully acknowledges the
support of a Compton GRO Fellowship, NASA grant 5-2687.
\bigskip

\centerline{REFERENCES}
\medskip

\ref Bildsten, L., Salpeter, E. E., \& Wasserman, I. 1992, ApJ,
384, 183

\ref Briggs, M. S. et al. 1996, IAU Circ. No. 6290

\ref Finger, M. H. et al. 1996a, IAU Circ. No. 6285

\ref Finger, M. H. et al. 1996b, IAU Circ. No. 6286

\ref Fishman, G. J., \& Meegan, C. A. 1995, ARA\&A, 33, 415

\ref Fishman, G. J. et al. 1995 IAU Circ. No. 6272

\ref Fishman, G. J. et al. 1996 IAU Circ. No. 6290

\ref Fushiki, I., \& Lamb, D. Q. 1987, ApJ, 323, L55

\ref Ghosh, P., \& Lamb F. K. 1979, ApJ 232, 259

\ref Hurley, K. in Proceedings of the Huntsville Symposium on
Gamma-Ray Bursts, AIP Conf. Proc., ed. C. Kouveliotou, M. S. Briggs,
and G. J. Fishman (AIP: New York in press).

\ref Illarionov, A. F., \& Sunyaev, R. A. 1979, A\&A, 39, 185

\ref Joss, P. C., \& Li, F. K. 1980, ApJ, 238, 287

\ref Joss, P. C. \& Rappaport, S. A. 1983, Nature, 304, 419

\ref Joss, P. C., Rappaport, S. \& Lewis, W. 1987, ApJ, 319, 180

\ref Kouveliotou, C. et al. 1996, IAU Circ. No. 6286

\ref Lewin, W. H. G., van Paradijs, J., \& Taam, R. E. 1993,
Space Sci. Rev., 62, 223

\ref M\'esz\'aros, P. 1992, {\it High-Energy Radiation from Magnetized
Neutron stars} (Chicago, University of Chicago Press).

\ref Miller, M. C. 1995, ApJ, 448, L29

\ref Miller, M. C. et al. 1996, IAU Circ. No. 6293

\ref Paciesas, W. S. et al. 1996, IAU Circ. No. 6284

\ref Schoelkopf, R. J., \& Kelley, R. L. 1991, ApJ, 375, 696

\ref Srinivasan, G., Bhattacharya, D., Muslimov, A. G., and Tsygan, A.
I. 1990, Current Science, 59, 31

\ref Stella, L., White, N. E., \& Rosner, R. 1986, ApJ, 308, 669

\ref Swank, J. et al. 1996, IAU Circ. No. 6291

\ref Taam, R. E. 1983, ApJ, 270, 694

\ref Taam, R. E., Woosley, S. E., Weaver, T. A., \& Lamb, D. Q. 1993,
ApJ, 413, 324

\ref Taam, R. E., Woosley, S. E., \& Lamb, D. Q. 1996, ApJ, in press

\ref Webbink, R. F., Rappaport, S. \& Savonije, G. J. 1983, ApJ, 
270, 678

\ref Zahn, J.-P. 1977, A\&A, 57, 583; erratum 67, 162 (1978)

\vfill\eject

\centerline{TABLE 1} 
\medskip
\centerline{\pcap Calculated Model Binary Parameters}
\medskip
\hrule
\vskip 2pt
\hrule
\vskip 3pt
\settabs 3 \columns
\+Binary parameters&If beginning Roche&If ending Roche\cr
\+&lobe overflow&lobe overflow\cr
\vskip 5pt
\hrule
\medskip
\+$M_c(M_\odot)$&1.0&0.20\cr
\+$M_{\rm ns}(M_\odot)$&1.4&2.0\cr
\+$M_{\rm He}(M_\odot)$&0.23&0.20\cr
\+$R_c(R_\odot)$&10.1&5.9\cr
\+$L_c(L_\odot)$&29.3&12.7\cr
\+$a(R_\odot)$&28.9&28.1\cr
\+$P_b$ (days)&11.76&11.76\cr
\+$f(M) \, (M_\odot)$&$1.31\times 10^{-4}$&$1.31\times 10^{-4}$\cr
\+$i$ (deg)&5.2&25\cr
\+Prob($i$)&$4\times 10^{-3}$&0.10\cr
\medskip
\hrule
\vskip 5pt

\noindent
$a$, total orbital separation; $f(M)$, mass function; Prob($i$), 
a priori probability that the orbital inclination is $\le i$; other
parameters defined in the text.

\vfill\eject\end

, Jean M. Quashnock

What can be said about the nature of the object that produces the X-ray
bursts and the persistent X-ray emission?  The existence of coherent
pulsations at a frequency $f = 2.14$ Hz, interpreted as rotation of the
object at a period $P = 0.467$ s, provides compelling evidence that
the object is not a white dwarf.  The regularity and coherence of the
pulsations argues strongly that the object is not a black hole, but a
strongly magnetic neutron star.

The main sequence star -- neutron star binary may itself have arisen
initially from a wide binary consisting of a main sequence star of $0.8
- 1.1 M_\sun$ and a main sequence star of, say, $15 M_\sun$.  Evolution
of the more massive star may have led to a common envelope phase
(Paczy\'nski 1983; Hofmeister \& Meyer-Hofmeister 1979), during which
the more massive star lost its hydrogen-rich envelope and the binary
separation markedly decreased, that resulted in a close binary
consisting of the main sequence star of $0.8 - 1.1 M_\sun$ and the $3 -
4 M_\sun$ degenerate helium-rich core of the more massive star.  The
eventual collapse of the degenerate helium-rich core and the subsequent
supernova explosion would produce the neutron star and a highly
eccentric binary with an orbital period $P_b \simgreat$ a few days.

Binary systems like GRO J1744-28 are thought to be the progenitors of
wide-binary pulsars (Joss \& Rappaport 1983, Paczy\'nski 1983, Savonije
1983).  Once the hydrogen-rich envelope of the giant has been
transferred onto the neutron star (or lost from the system), the giant
shrinks within its Roche lobe and mass transfer ceases, leaving behind
a wide binary consisting of the degenerate helium-rich core of the
giant and the neutron star.  After mass transfer ceases, the binary
orbital period is virtually fixed, since the stars are too widely
separated for tidal dissipation or gravitational radiation to alter the
orbital period.  GRO J1744-28 can therefore be expected to evolve
eventually into a system that is similar in some ways, but not others,
to observed wide-binary pulsars (Phinney 1992).

The present binary orbital period of GRO J1744-28, which will be its
final orbital period if, as seems highly likely, the companion has
nearly exhausted its hydrogen-rich envelope, is similar to those of PSR
1855+09 and PSR 2303+46.  The small mass of the companion star is
similar to that in PSR 1855+09 but not PSR 2303+46 (which has a
companion star mass $M_c = 1.5 M_\sun$ and probably a different
evolutionary history).  On the other hand, the dipole magnetic field of
the neutron star in GRO J1744-28 is larger, but perhaps not
qualitatively so, than that of PSR 2303+46 (which has a dipole magnetic
field $B_{\rm dipole} \approx 7 \times 10^{11}$ G) but is much larger
than that of PSR 1855+09 (which has a dipole magnetic field $B_{\rm
dipole} \approx 3 \times 10^8$ G).

The bursting behavior of the neutron star in this system is similar in
some ways, but not in others, to each of the three known classes of
burst sources; X-ray burst sources,$^{19}$ soft $\gamma$-ray
repeaters,$^{20}$ and $\gamma$-ray burst sources.$^{21}$  The durations
$t_{\rm dur}$ of the bursts and the existence of persistent X-ray
emission are similar to X-ray burst sources, but the characteristic
spectral energy $\langle E \rangle$ of the bursts and the burst rate
$r_b$ are a factor $\approx$ 20 higher.  Both $\langle E \rangle$ and
$r_b$ are similar to some SGRs, but $t_{\rm dur}$ is $\approx$ 100
times longer and no persistent X-ray emission is seen from SGRs.  Only
$t_{\rm dur}$ is similar to that of $\gamma$-ray bursts, while $\langle
E \rangle$ is a factor of 10 $-$ 50 smaller, repeating of $\gamma$-ray
bursts, while widely discussed, has not been conclusively demonstrated,
and no persistent X-ray emission is seen from $\gamma$-ray burst
sources.  Both $t_{\rm dur}$ and $r_b$ are similar to some bursts from
the singular source called the Rapid Burster,$^5$ as is the existence
of persistent X-ray emission, but $\langle E \rangle$ is not.

\centerline{TABLE 2} 
\medskip
\centerline{\pcap Comparison of Bursts from GRO J1744-28}
\centerline{\pcap with Bursts from Other Types of Burst Sources}
\medskip
\hrule
\vskip 2pt
\hrule
\vskip 3pt
\settabs 6 \columns
\+Property&X-ray&SGR&$\gamma$-ray&Rapid&GRO J1744-28\cr
\+\ \ of bursts&bursts&bursts&bursts&Burster\cr
\vskip 5pt
\hrule
\medskip
\+$t_{dur}$ (s)&3-1000&0.2-3&$10^{-3}-10^3$&2-600&8 to $>$30\cr
\+$r_b$ (hr$^{-1}$)&$10^{-2}-10$&$10^{-4}-10$&Unknown&$1-400$&
$1.6-18$\cr
\+$\langle E\rangle$ (keV)&6&20&$10^2-10^3$&6&14\cr
\+Persistent\cr
\+\ \ X-ray em.&Yes&Yes&No&Yes&Yes\cr
\+$F_b^{\rm pk}/F_p$&$2-10^2$&$10^5-10^7$&NA&$10$&$2-10$\cr
\medskip
\hrule

(4)  Observation of some spectral change during the bursts,
particularly the disappearance of a softer emission component, would be
supportive of the thermonuclear flash model.

(5) The burst spectra may exhibit an additional, softer emission
component than the spectrum of the persistent emission, due to leakage
of photons from around the accreting area at the magnetic polar cap
during the burst.  However, this softer component is expected to be
only a few percent of the total burst luminosity.

\ref Lyubarski, Yu. E., \& Sunyaev, R. A. 1982, Sov. Astr. J.
Lett., 8, 612

\ref Meyer, F. \& Meyer-Hofmeister, E. 1979, A\&A, 78, 167

\ref Paczy\`nski, B. 1976, {\it Structure and Evolution of Close
Binary Systems}, IAU Symp. No. 88, 75

\ref Paczy\'nski, B. 1983, Nature, 304, 421

\ref Phinney, S. 1992, Phil. Trans. R. Soc. Lond. A, 341, 39

\ref Savonije, G. J. 1983, Nature, 304, 422

\ref Spruit, H. C., \& Taam, R. E. 1986, ApJ, 402, 593

\ref Wang, Y.-M., \& Frank, J. 1981, A\&A, 93, 255